%% file: main.tex
\documentclass[10pt,conference]{IEEEtran}
\usepackage[T1]{fontenc}
\usepackage{booktabs}
\usepackage{graphicx}
\usepackage{xcolor}
\usepackage{amsmath}
\usepackage{microtype}
\usepackage[caption=false,font=footnotesize]{subfig}
\usepackage{tikz}
\usetikzlibrary{arrows.meta,calc,positioning,fit,backgrounds}
\usepackage[hidelinks]{hyperref}
\hypersetup{pdftitle={SWHForest: A Structural Curation and Search Layer for Large-Scale Code Corpora},
  pdfauthor={Federico Pennino, Andrea Gurioli, Maurizio Gabbrielli, Paolo Ferragina}}

\input{numbers}

\newcommand{\tool}{\textsc{SWHForest}}
\newcommand{\code}[1]{\texttt{\small #1}}
\newcommand{\tcode}[1]{\texttt{\scriptsize #1}}

\begin{document}

\title{SWHForest: A Structural Curation and Search Layer\\for Large-Scale Code Corpora}

\author{\IEEEauthorblockN{Federico Pennino$^\dag$, Andrea Gurioli$^\dag$, Maurizio Gabbrielli$^\dag$, and Paolo Ferragina$^\ddag$}
\IEEEauthorblockA{$^\dag$Department of Computer Science and Engineering, University of Bologna, Italy\\
\{federico.pennino2, andrea.gurioli5, maurizio.gabbrielli\}@unibo.it\\
$^\ddag$Sant'Anna School of Advanced Studies, Pisa, Italy \\ paolo.ferragina@santannapisa.it}}

\maketitle

\begin{abstract}
Pre-training code language models relies on massive code corpora usually streamed over the internet. However, researchers rarely need a multi-terabyte archive in its entirety; instead, their workflow centers on \emph{curated subsets}---selecting well-documented files, filtering out repetitive boilerplate, auditing for benchmark contamination, or retrieving specific code patterns. Yet, because modern corpora are typically published as file pointers and basic attributes rather than source text, \emph{no general structural layer exists to select which files to download}: curating a subset requires downloading and parsing millions of files first just to discard most of them. We present \tool{}, an open-source framework and public dataset that provides this missing structural layer for any pointer-based corpus without redistributing source code. A single pass over the archive parses each file into an AST and immediately discards the source text, keeping only its structural summary. Queried locally or exposed by archive providers as a server-side filtering API, \tool{} selects matching files first---via SQL or index intersection---and downloads only surviving candidates. Across the ten largest languages of Stack-Edu (\NRelBlobs{}M unique files, \NRelDefs{}M definitions), \tool{} answers curation queries in milliseconds while cutting downloads by \NRelCurAllSaved{}\%--\NRelCurSaved{}\%, reducing live structural searches from hours of per-language streaming to seconds or minutes.
\end{abstract}

\section{Introduction}
\label{sec:intro}

Modern code language models are commonly trained on large code corpora collected from publicly available sources on the internet~\cite{kocetkov2023stack, lozhkov2024starcoder2}. Increasingly, large-scale datasets decouple \emph{corpus metadata} from \emph{raw file storage}: rather than shipping multi-terabyte archives of source text (as in The Stack~v1~\cite{kocetkov2023stack}), modern releases such as The Stack~v2~\cite{lozhkov2024starcoder2} and Stack-Edu~\cite{benallal2025smollm2} distribute lightweight tables of content-addressed file identifiers (\code{blob\_id}) paired with basic metadata (repository, path, byte length, license, quality score), while actual file contents (\emph{blobs}) reside in remote object storage such as the Software Heritage Archive (SWH)~\cite{dicosmo2017swh} or cloud buckets. Distributing pointers rather than source text avoids duplicating terabytes of code while respecting upstream opt-outs and preserving historical provenance~\cite{dicosmo2020referencing,katzy2024license,jahanshahi2025cracks}.

\begin{figure}[t]
\centering
\subfloat[Building the structural layer (one-time pass; zero source code stored)]{\begin{tikzpicture}[x=1mm,y=1mm,font=\scriptsize,
  >={Stealth[length=1.5mm]}, line width=0.5pt,
  box/.style={draw=black!60, rounded corners=1.8pt, align=center, inner sep=2pt, minimum height=12.5mm},
  ext/.style={box, fill=black!4},
  ours/.style={box, draw=blue!65!black, fill=blue!6},
  tab/.style={draw=blue!55!black, fill=white, rounded corners=1.2pt, align=left,
              inner sep=1.8pt, text width=79.5mm, font=\scriptsize, anchor=north west},
  note/.style={font=\scriptsize, inner sep=1pt, text=black!75}]

\node[ext, text width=23.5mm] (edu) at (12.5, 0)
  {\textbf{1. Corpus IDs}\\[0.2ex]
   File IDs (\texttt{blob\_id})\\[-0.1ex]+ metadata (\emph{no code})};
\node[ours, text width=23.5mm] (sel) at (42.5, 0)
  {\textbf{2. Hash Sample}\\[0.2ex]
   SHA-1 prefix filter\\[-0.1ex](\textbf{52.1M} unique files)};
\node[ours, text width=25.5mm] (prs) at (73.5, 0)
  {\textbf{3. Stream \& Parse}\\[0.2ex]
   Stream remote blobs;\\[-0.1ex]Tree-sitter discards text};

\draw[->] (edu) -- (sel);
\draw[->] (sel) -- (prs);

\node[anchor=north west, inner sep=0.5pt, font=\scriptsize\bfseries, text=blue!60!black]
  (relhdr) at (2.0, -11.5) {Precomputed Structural Layer (37.7\,GB Parquet, 0\,B source code)};
\node[tab, below=0.9mm of relhdr.south west, anchor=north west] (cat)
  {\textbf{\texttt{catalog.parquet}:} 17 syntactic metrics per file (functions, classes, docstrings, depth)};
\node[tab, below=0.9mm of cat.south west, anchor=north west] (fun)
  {\textbf{\texttt{functions} \& \texttt{shapes}:} byte spans + rename-invariant AST fingerprints (\textbf{341M} defs)};
\node[tab, below=0.9mm of fun.south west, anchor=north west] (idx)
  {\textbf{\texttt{structure-*.parquet}:} search index mapping AST building blocks $\to$ \textbf{file ID lists}};
\begin{scope}[on background layer]
\node[draw=blue!65!black, fill=blue!7, rounded corners=2pt, line width=0.7pt,
      inner sep=1.8pt, fit=(relhdr)(cat)(idx)] (rel) {};
\end{scope}

\draw[->] (prs.south) -- (prs.south |- rel.north);
\end{tikzpicture}\label{fig:build}}\\[1.8mm]
\subfloat[Curating and querying locally before downloading files]{\begin{tikzpicture}[x=1mm,y=1mm,font=\scriptsize,
  >={Stealth[length=1.5mm]}, line width=0.5pt,
  box/.style={draw=black!60, rounded corners=1.8pt, align=left, inner sep=2pt},
  ours/.style={box, draw=blue!65!black, fill=blue!6},
  qry/.style={box, draw=orange!75!black, fill=orange!8},
  outb/.style={box, draw=green!50!black, fill=green!7}]

\node[qry, text width=21.5mm, minimum height=17.5mm] (pat) at (11.5, 0)
  {\textbf{1. Code Pattern}\\[0.3ex]
   Find functions:\\[0.1ex]
   \texttt{def train():}\\[-0.2ex]
   \texttt{\ \ ...\ .fit()}};
\node[ours, text width=28.5mm, minimum height=17.5mm] (fea) at (41.0, 0)
  {\textbf{2. Break into Pieces}\\[0.3ex]
   Every match must have:\\[-0.1ex]
   $\bullet$ \texttt{def} named \texttt{"train"}\\[-0.1ex]
   $\bullet$ call to \texttt{".fit"}\\[-0.1ex]
   $\bullet$ \texttt{def} enclosing a call};
\node[ours, text width=25.0mm, minimum height=17.5mm] (pos) at (72.5, 0)
  {\textbf{3. Intersect File IDs}\\[0.3ex]
   Keep only files present\\[-0.1ex]
   in \textbf{all} piece lists\\[-0.1ex]
   (\textbf{skips 99.5\% median})};

\draw[->] (pat) -- (fea);
\draw[->] (fea) -- (pos);

\node[outb, text width=39.0mm, minimum height=13.5mm] (ex) at (20.2, -20.2)
  {\textbf{If Pattern Has 1 Piece (Exact)}\\[0.2ex]
   e.g.\ \texttt{except\{pass\}} or \texttt{eval()}\\[-0.1ex]
   Index list is exact $\to$ \textbf{0 files fetched}};
\node[qry, text width=39.0mm, minimum height=13.5mm] (ver) at (65.2, -20.2)
  {\textbf{If Pattern Combines Pieces}\\[0.2ex]
   e.g.\ \texttt{def train():\ ...\ .fit()}\\[-0.1ex]
   Fetch \textbf{0.53\% median} \& verify (\texttt{ast-grep})};

\coordinate (branch) at ($(pos.south)+(0,-2.3)$);
\draw[->] (pos.south) -- (branch) -| (ex.north);
\draw[->] (pos.south) -- (branch) -| (ver.north);
\end{tikzpicture}\label{fig:query}}
\caption{Architecture and workflow of \tool{}. (a)~Our pipeline streams and parses remote files once (instantiated on 52.1M Stack-Edu/SWH files), releasing compact Parquet tables and a structural index with zero source code. (b)~Users filter and query the local index first, downloading only matching candidate files.}
\label{fig:pipeline}
\end{figure}
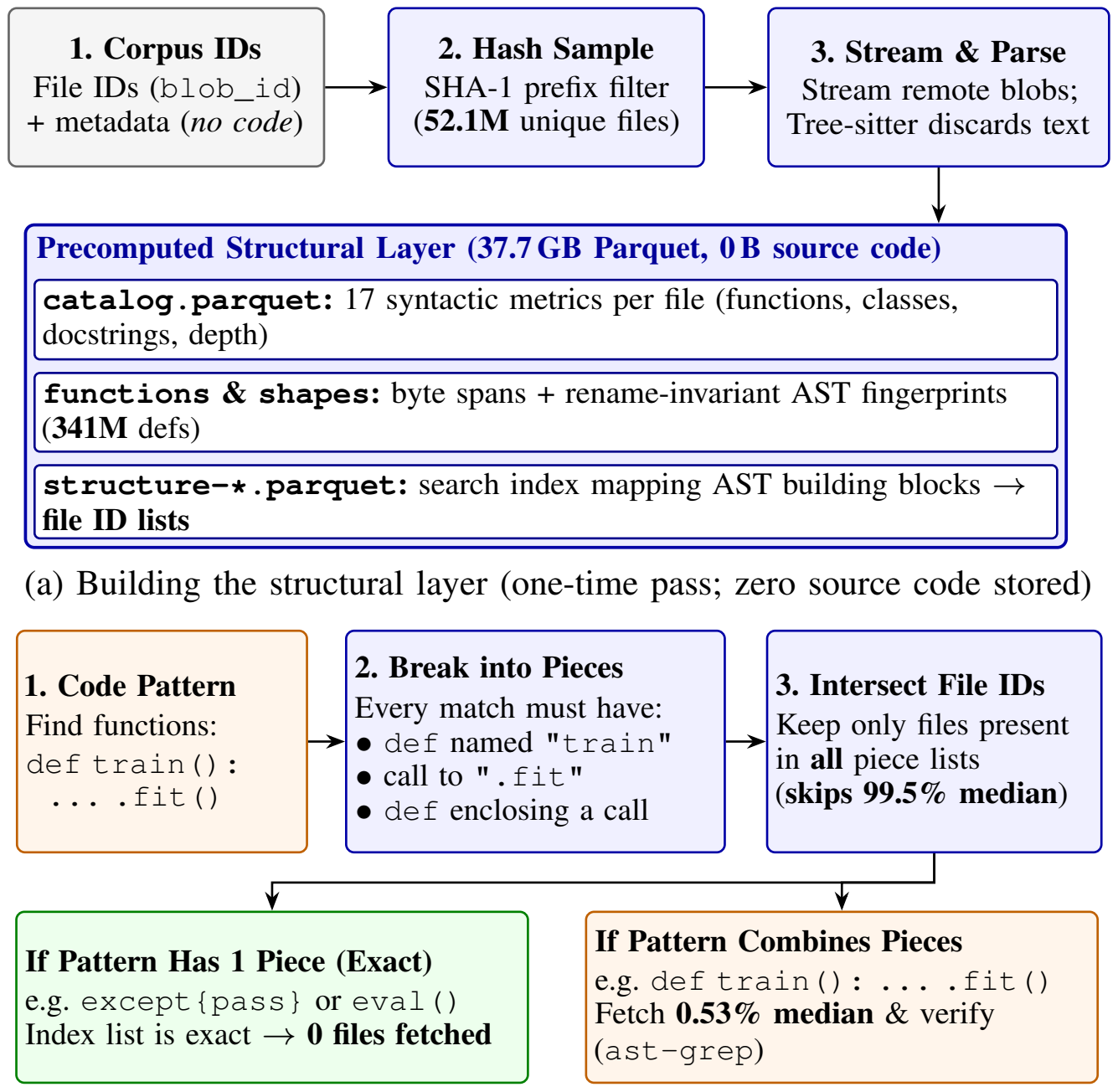

However, when the goal is \emph{data curation or analytics} rather than downloading the entire archive, pointer-only distribution creates a severe bottleneck: \emph{pointer-based corpora lack a structural layer to select files by content and syntax before downloading them}. Suppose a team wants to filter for documented files with at least five definitions, strip out repetitive Java/PHP getters, audit for benchmark contamination, or locate specific constructs (e.g., \code{defer} inside a Go loop or \code{transmute} inside Rust \code{unsafe}). Because the metadata table records only byte size, answering any of these questions requires downloading and parsing millions of remote files: even on a 128-core server, streaming and parsing \NRelBlobs{}M files takes \NRelHours{} hours. Existing infrastructures such as World of Code~\cite{ma2019woc} and the SWH Graph~\cite{pietri2019swhgraph,boldi2020swhgraph} index commit histories and provenance rather than intra-file syntax, while query platforms such as Boa~\cite{dyer2013boa} and CodeQL~\cite{avgustinov2016ql} must first ingest raw source code into custom databases.

We designed \tool{} to provide this missing layer for any pointer-based or remote code corpus. Keyed by \code{blob\_id}, its precomputed Parquet tables and inverted AST index join directly with upstream metadata, letting users filter files via SQL or intersect AST building-block lists locally and fetch only surviving candidates. Our contributions are:
\begin{itemize}\setlength{\itemsep}{0.5pt}
\item \textbf{A Storage-Agnostic Structural Layer and Tool:} An open-source framework and compact AST feature index (integrated with \code{ast-grep}~\cite{astgrep} across ten languages) that resolves single-constraint patterns from the local index without downloading files and prunes complex AST searches to a \NMLCandMedPct{}\% median with zero false negatives.
\item \textbf{A 52.1M-File Multi-Language Public Catalog:} A public dataset covering a uniform 40\% sample of the ten largest languages in Stack-Edu (\NRelBlobs{}M files across \NRelReposM{}M repository--language pairs), with per-file syntactic metrics, byte spans for \NRelDefs{}M definitions (functions, methods, classes), and 64-bit structural \emph{shape} hashes.
\item \textbf{Reproducible Views and Ten-Language Evaluation:} A portable JSON manifest format for citing and rebuilding filtered subsets, evaluated across all ten languages on SQL curation, boilerplate/contamination detection, and structural search.\footnote{Code: \url{https://github.com/halykoss/swhforest}; data: \url{https://huggingface.co/datasets/halykoss/swhforest} (Zenodo DOI pending).}
\end{itemize}

\section{System Overview and Data Model}
\label{sec:model}

Fig.~\ref{fig:pipeline} shows how \tool{} decouples structural curation from raw file storage. The framework requires only a content hash (\code{blob\_id}) per file and a pluggable reader for local disks, cloud buckets, or SWH. At build time (Fig.~\ref{fig:build}), the pipeline reads file IDs from the manifest, streams each file through Tree-sitter~\cite{treesitter}, and discards the source text after writing its syntactic metrics, definition fingerprints (\code{shape}), and AST feature posting lists into Parquet (\code{catalog}, \code{functions}, \code{shapes}, \code{structure-*}). At query time (Fig.~\ref{fig:query}), users either download the tables and index for their target languages locally (37.7\,GB in total; Table~\ref{tab:release}) or, when hosted by an archive provider, query a filtering API that returns matching \code{blob\_id}s directly, fetching only the surviving source files.

\textit{Dataset Tables.} Each language release (Table~\ref{tab:release}) comprises five Parquet tables linked by \code{blob\_id} (the 40-hex Git SHA-1):
\begin{itemize}\setlength{\itemsep}{0.5pt}
\item \code{rows.parquet} maps each unique \code{blob\_id} in the sample to a compact integer file ID (\code{row}) and upstream metadata.
\item \code{catalog.parquet} has one row per file, joining representative upstream metadata (repository, path, license, score) with 17 syntactic metrics: size (bytes, lines), tree structure (AST nodes, leaves, depth, syntax errors), documentation (comments, docstrings, comment ratio), and code elements (definitions, distinct shapes, imports, strings, identifiers).
\item \code{functions.parquet} lists all \NRelDefs{}M definitions (\code{function}, \code{method}, or \code{class}) with their \code{blob\_id}, byte offsets, AST node count, docstring flag, and 64-bit \code{shape} (Section~\ref{sec:data}), allowing functions to be sliced from fetched files without re-parsing.
\item \code{shapes.parquet} groups definitions sharing an identical \code{shape} within a language, recording cluster size, mean AST size, and a sample byte span.
\item \code{structure-<k>.parquet} stores the inverted structural index (in 2M-file shards), mapping each AST feature to the sorted list of file IDs containing it.
\end{itemize}

\textit{Reproducible Views.} To share curated subsets without redistributing code, \tool{} exports a \emph{view}: a JSON manifest recording the filter query, tool and grammar versions, sorted \code{blob\_id} list, and SHA-256 checksum. Each language ships \code{view\_release.json} for the 40\% sample; custom views are cited by checksum and re-fetched on demand.

\begin{table}[t]
\centering
\caption{Our public \tool{} release across the ten largest languages of Stack-Edu (40\% uniform sample of 200\,B--200\,kB files). \textbf{Files}: unique files (\texttt{blob\_id}); \textbf{Defs}: functions, methods, and classes; \textbf{Docs}: files with doc comments (Go has no distinct syntax); \textbf{Err}: files with syntax error nodes (mainly unexpanded C/C++ macros); \textbf{Tables}/\textbf{Index}: disk sizes.}
\label{tab:release}
\setlength{\tabcolsep}{3.2pt}
\footnotesize
\begin{tabular}{@{}lrrrrrr@{}}
\toprule
Language & Files & Defs & Docs & Err & Tables & Index \\
 & (M) & (M) & (\%) & (\%) & (GB) & (GB) \\
\midrule
Java & \NRelJavaBlobs & \NRelJavaDefs & \NRelJavaDoc & \NRelJavaErr & \NRelJavaTab & \NRelJavaIdx \\
Python & \NRelPyBlobs & \NRelPyDefs & \NRelPyDoc & \NRelPyErr & \NRelPyTab & \NRelPyIdx \\
C++ & \NRelCppBlobs & \NRelCppDefs & \NRelCppDoc & \NRelCppErr & \NRelCppTab & \NRelCppIdx \\
JavaScript & \NRelJsBlobs & \NRelJsDefs & \NRelJsDoc & \NRelJsErr & \NRelJsTab & \NRelJsIdx \\
C\# & \NRelCsBlobs & \NRelCsDefs & \NRelCsDoc & \NRelCsErr & \NRelCsTab & \NRelCsIdx \\
PHP & \NRelPhpBlobs & \NRelPhpDefs & \NRelPhpDoc & \NRelPhpErr & \NRelPhpTab & \NRelPhpIdx \\
C & \NRelCBlobs & \NRelCDefs & \NRelCDoc & \NRelCErr & \NRelCTab & \NRelCIdx \\
TypeScript & \NRelTsBlobs & \NRelTsDefs & \NRelTsDoc & \NRelTsErr & \NRelTsTab & \NRelTsIdx \\
Go & \NRelGoBlobs & \NRelGoDefs & -- & \NRelGoErr & \NRelGoTab & \NRelGoIdx \\
Rust & \NRelRustBlobs & \NRelRustDefs & \NRelRustDoc & \NRelRustErr & \NRelRustTab & \NRelRustIdx \\
\midrule
Total & \NRelBlobs & \NRelDefs & & & \NRelTabGB & \NRelIdxGB \\
\bottomrule
\end{tabular}
\end{table}

\section{Structural Indexing and Public Release}
\label{sec:data}

\textit{Rename-Invariant Structural Fingerprints (\code{shape}).} To detect repeated code templates and renamed clones without storing source text, \tool{} strips all identifier names and literal values (ignoring identifier reuse) from each definition's syntax tree, keeping only grammar node types (e.g., \code{identifier} or \code{return}). It then hashes the subtree bottom-up into a 64-bit Merkle fingerprint~\cite{merkle1987signature} from each node's grammar type and child hashes, storing the root hash as \code{shape} in \code{functions.parquet} and \code{shapes.parquet}. Thus, \code{def get\_x(self): return self.x} and \mbox{\code{def get\_y(o): return o.y}} share a \code{shape} (matching Type-2 clones~\cite{roy2009comparison}, retaining comments).

\textit{Structural Search Index.} Text search engines often rely on trigram indexes to match literal substrings~\cite{cox2012trigram,clem2023blackbird}, which cannot filter purely structural patterns such as nested loops or empty exception handlers. At the other extreme, merging shared subtrees into a deduplicated AST DAG~\cite{buneman2003compressed,ledilavrec2022hyperast} on a 100k-file sample required \NForestRatio{}$\times$ the compressed text size (vs.\ $\sim$0.4$\times$ for our index) and kept every literal token. Instead, drawing on feature indexes for tree and graph databases~\cite{goldman1997dataguides,yan2004gindex,zhang2006fix}, \tool{} decomposes each file's AST into three simple building blocks (Fig.~\ref{fig:query}):
\begin{enumerate}\setlength{\itemsep}{0.5pt}
\item \textbf{Parent--child links (\code{E}):} a grammar construct and the syntactic role of its direct child (e.g., an \code{if} statement and its \code{condition} or \code{body}, ignoring punctuation);
\item \textbf{Literals in context (\code{T}):} a literal token with its enclosing construct and role (e.g., \code{"eval"} as the target of a \code{call});
\item \textbf{Nesting pairs (\code{A}):} an ancestor and a descendant construct anywhere inside it (e.g., a \code{for} loop enclosing \code{await}).
\end{enumerate}
In \code{structure-<k>.parquet}, each building block is hashed to a 64-bit key and mapped to a sorted list of file IDs. A query combines node types, field roles, literals, and nesting constraints, translating directly to an \code{ast-grep}~\cite{astgrep} relational rule. \tool{} extracts only the building blocks required by \emph{every} match and intersects their file lists (rarest first). Because every match must contain all extracted blocks (and hash collisions only add candidates), pruning never drops a match (\emph{zero false negatives}). When a query reduces to a \emph{single} block---one nesting pair (e.g., \code{except} enclosing \code{pass} at any depth) or one literal in a role (e.g., calling \code{forName} on any receiver)---its posting list gives the exact set of matching files (modulo 64-bit collisions, none observed) with zero blobs fetched; for multi-block queries, list intersection finds all files where those blocks co-occur and \code{ast-grep} verifies that they form a match. We verified zero missed matches across 1.2M random pattern--file pairs in all ten grammars, 140k cached files, and all \NMLSoundN{} benchmark queries.

\textit{Public Release Construction and Validation.} Although our pipeline runs on any local or remote corpus, we release a public catalog for Stack-Edu~\cite{benallal2025smollm2} (revision \code{eeec5ca}), an educational subset of The Stack~v2~\cite{lozhkov2024starcoder2}. We cover its ten largest languages (Table~\ref{tab:release}), excluding Markdown, SQL, Shell, and the two smallest (Ruby, Swift). Among deduplicated files of 200\,B--200\,kB, we select all files whose Git SHA-1 (\code{blob\_id}) starts below \code{66666666} (the lowest 40\% of the hash space). Because SHA-1 hashes are uniform, this yields a deterministic, seed-free 40\% sample per language (\NRelSelPct{}\% of eligible files; \NRelBlobsExact{} total) verifiable from the ID and subsuming smaller prefix samples (e.g., 10\% below \code{19999999}). Streaming and parsing all \NRelBlobs{}M files from SWH with Tree-sitter~0.26.0~\cite{treesitter} on a 128-core node took \NRelHours{}\,h ($\sim$\NRelRate{} files/s; \NRelMissing{} unavailable blobs).

\begin{figure}[t]
\centering
\includegraphics[width=\columnwidth]{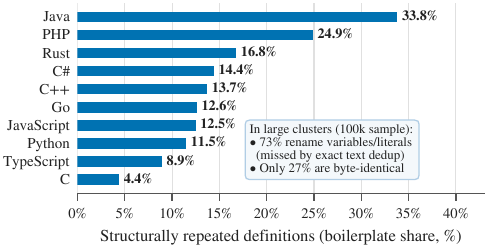}
\caption{Percentage of non-trivial definitions ($\ge 20$ AST nodes) in each language whose exact syntax-tree structure (\texttt{shape}) repeats more than 100 times across the corpus. In our 100k-file Python/JS sample, 73\% of these repeated definitions rename identifiers or literals and are missed by exact-text deduplication.}
\label{fig:boiler}
\end{figure}

Per-language JSON specifications map Tree-sitter node types to functions, methods, classes, comments, and docstrings (counting lambdas/closures as functions in all languages except Python and C). On 50k Python files, our counts matched CPython's \code{ast} module on \NValAgree{}\% of the \NValChecked{} doubly-parsed files (\NValSkipPct{}\% rejected by one parser). Across the other nine languages, auditing 110 random definitions and 300 files per language confirmed 100\% precision and 99.6\% recall (7 missed definitions in 2,700 files, 5 in a PHP file without \code{<?php}).

\section{Applications and Use Cases}
\label{sec:uses}

\textit{Use Case 1: Millisecond Corpus Curation via SQL.} Selecting files across our ten-language release (\NRelBlobs{}M rows) that exceed 10\,kB, include a docstring, and define $\ge 5$ functions/methods takes one SQL query over \code{catalog.parquet}. DuckDB executes it in \textbf{\NRelCurAllMs{}\,ms}, returning \NRelCurAllMatchesM{}M files (\NRelCurAllGB{}\,GB of source text; \textbf{\NRelCurMs{}\,ms} for \NRelCurMatches{} Python files). Without \tool{}, size-only filtering leaves \NRelSizeOnlyAllM{}M files (\NRelSizeOnlyAllGB{}\,GB) to download and parse; \tool{} cuts download volume by \textbf{\NRelCurAllSaved{}\%} (\NRelCurSaved{}\% on Python) without fetching or parsing rejected files.

\textit{Use Case 2: Boilerplate and Contamination Detection.} Although The Stack~v2 deduplicates files upstream, heavy repetition persists across definitions in different files. Using \code{shapes.parquet}, finding definitions with identical AST structure up to renaming is an instant lookup. Fig.~\ref{fig:boiler} shows the share of non-trivial definitions ($\ge 20$ AST nodes) whose shape repeats $>100$ times: from \NRelCBoiler{}\% in C to \NRelPhpBoiler{}\% in PHP and \NRelJavaBoiler{}\% in Java, dominated by accessors and constructors. On a 100k-file Python/JS sample, \textbf{only \NExactPct{}\% of definitions in these large clusters are byte-identical} (Type-1); the remaining \textbf{73\% differ in identifiers or literals}, and \textbf{88\%} of these have token 5-gram Jaccard $<0.7$ (median $0.32$) to every cluster peer, escaping MinHash. In a manual audit of 60 non-identical pairs, all are recurring boilerplate templates (constructors, accessors, or wrapper calls) and \textbf{82\%} call the exact same API with renamed variables (Type-2 clones). The same lookup also serves as a fast prefilter for benchmark contamination~\cite{allamanis2019adverse,riddell2024contamination}: matching reference solutions against our Python release flags \textbf{\NHumanEvalHits{}/164 HumanEval}~\cite{chen2021humaneval} and \textbf{\NMbppHits{}/496 MBPP}~\cite{austin2021mbpp} tasks. Manual inspection confirms that \textbf{10 of 15} HumanEval hits and \textbf{39 of 50} sampled MBPP hits implement the same task logic (the rest are generic one-liners), and that most multi-statement matches again have token Jaccard $<0.7$.

\begin{figure}[t]
\centering
\includegraphics[width=\columnwidth]{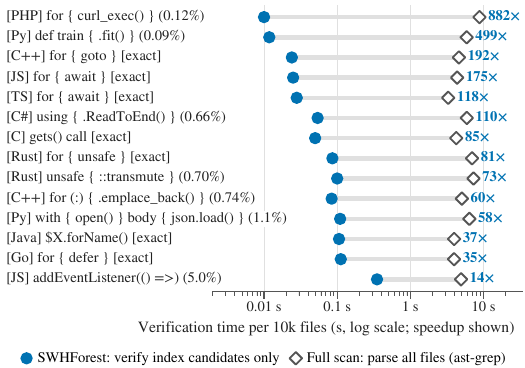}
\caption{Execution time for 14 representative structural queries out of \NMLSoundN{} across all 10 languages (cached locally on one core; median of 3--5 runs), comparing a full scan with \texttt{ast-grep} against \tool{} index pruning followed by \texttt{ast-grep} verification on candidate files. Labels show the candidate percentage (or \texttt{[exact]} when answered directly by the index) and speedup; \texttt{X \{ Y \}}: \texttt{Y} at any depth in \texttt{X}.}
\label{fig:queries}
\end{figure}

\textit{Use Case 3: Pattern and Security-Rule Search.} Studies of API misuse, concurrency bugs, or training subsets need all files matching a pattern (e.g., \code{defer} in a Go loop, \code{transmute} in Rust \code{unsafe}) or a static-analysis rule. \tool{} compiles structural patterns into feature queries, intersects the posting lists locally, and optionally joins file IDs with \code{catalog.parquet} (license/size/doc filters) or \code{functions.parquet} (to slice enclosing functions). We translated \NSgN{} Semgrep~\cite{semgrep} rules in Python/JS into this fragment (dropping negation, metavariable equality, and taint). Against the translated rules, the index misses zero matches, resolves \NSgExact{} single-constraint rules directly without fetching blobs, and prunes multi-clause rules to a median of \textbf{\NSgFetchPct{}\%} of corpus bytes.

\section{Search Performance and Pruning Evaluation}
\label{sec:eval}

\textit{Local Verification Speedup and Baseline Comparison.} We benchmark structural search across \NMLSoundN{} queries spanning all ten languages, using the Rust-based \code{ast-grep} engine~\cite{astgrep} to verify candidate files after index pruning (Fig.~\ref{fig:queries}). On cached local files (50k files each for Python/JS and 5k for the other eight languages), filtering with \tool{} accelerates \code{ast-grep} verification by a geometric mean of \textbf{\NMLGeoSpeed{}$\times$} (\NRMinSpeed{}$\times$--\NMLMaxSpeed{}$\times$ across all languages; \NRGeoSpeed{}$\times$ warm and \NRGeoSpeedLoad{}$\times$ cold on Python/JS) with zero missed matches. Recording syntactic roles and nesting yields \textbf{\NTokRatio{}$\times$ fewer false-positive candidates} than a typed-token index \code{(leaf\_type, text)}, and \textbf{8.8$\times$ fewer false positives} on the 50 lexically anchored queries (\textbf{207$\times$} on the 11 purely structural queries, e.g., loops with \code{await} or \code{defer}) than a substring/trigram prefilter (tighter on 4 literal-heavy queries); pruning is least selective (3.1$\times$) when common constructs co-occur (e.g., \code{print} in \code{for}).

\textit{Live End-to-End Search over 52.1M Remote Files.} Across the full release, \NMLExactN{} of the \NMLSoundN{} queries are answered directly by the index with zero downloads (returning all matching \code{blob\_id}s), while the rest only need to download and verify \textbf{\NMLCandMedPct{}\%} of the files (median). Table~\ref{tab:e2e} shows end-to-end times for 12 queries across all ten languages live over the network: queries answered directly by the index (\code{exact}) finish in \textbf{<0.01--0.52\,s} with zero traffic, while the others download and verify the surviving files in \textbf{11\,s to 2.2\,min}. Even compared to streaming a single language at \NRelRate{}\,files/s (3.8\,min on Rust to 2.4\,h on Java, or \NRelHours{}\,h for all ten), index filtering is \textbf{4$\times$ to 171$\times$ faster}.

\section{Related Work}
\label{sec:related}

\begin{table}[t]
\centering
\caption{End-to-end search over the full 10-language release (\NRelBlobs{}M files): warm index lookup (\NRelIdxLoadS{}\,s cold load on Py/JS), live remote fetch (256 threads), and \texttt{ast-grep} verification (\texttt{exact}: exact matching file set from index). \texttt{X \{ Y \}}: \texttt{Y} at any depth in \texttt{X}.}
\label{tab:e2e}
\setlength{\tabcolsep}{2.3pt}
\footnotesize
\begin{tabular}{@{}llrrrr@{}}
\toprule
Lang. & Query Pattern & Cand. & Matches & Fetch & Total \\
\midrule
Py   & \tcode{except \{ pass \}} & \NEExceptPassCand & exact & 0\,s & \NEExceptPassTotal\,s \\
JS   & \tcode{for (..) \{ await \}} & \NEJsAwaitInLoopCand & exact & 0\,s & \NEJsAwaitInLoopTotal\,s \\
Java & \tcode{\$X.forName(..)} & \NEJavaFornameCand & exact & 0\,s & \NEJavaFornameTotal\,s \\
Rust & \tcode{unsafe \{ ::transmute \}} & \NERustUnsafeTransmuteCand & \NERustUnsafeTransmuteHits & \NERustUnsafeTransmuteFetch\,s & \NERustUnsafeTransmuteTotal\,s \\
PHP  & \tcode{for (..) \{ curl\_exec() \}} & \NEPhpForCurlCand & \NEPhpForCurlHits & \NEPhpForCurlFetch\,s & \NEPhpForCurlTotal\,s \\
TS   & \tcode{catch \{ console.error \}} & \NETsTryConsoleErrCand & \NETsTryConsoleErrHits & \NETsTryConsoleErrFetch\,s & \NETsTryConsoleErrTotal\,s \\
Py   & \tcode{def train():\ .. .fit()} & \NETrainFitCand & \NETrainFitHits & \NETrainFitFetch\,s & \NETrainFitTotal\,s \\
Java & \tcode{synchronized \{ .wait() \}} & \NEJavaSyncWaitCand & \NEJavaSyncWaitHits & \NEJavaSyncWaitFetch\,s & \NEJavaSyncWaitTotal\,s \\
C\#  & \tcode{using (..) \{ .ReadToEnd() \}} & \NECsUsingReadCand & \NECsUsingReadHits & \NECsUsingReadFetch\,s & \NECsUsingReadTotal\,s \\
C    & \tcode{for (..) \{ realloc(..) \}} & \NECForReallocCand & \NECForReallocHits & \NECForReallocFetch\,s & \NECForReallocTotal\,s \\
Go   & \tcode{defer \{ .Unlock() \}} & \NEGoDeferUnlockCand & \NEGoDeferUnlockHits & \NEGoDeferUnlockFetch\,s & \NEGoDeferUnlockTotal\,s \\
C++  & \tcode{for (:) \{ .emplace\_back() \}} & \NECppRangeEmplaceCand & \NECppRangeEmplaceHits & \NECppRangeEmplaceFetch\,s & \NECppRangeEmplaceTotal\,s \\
\bottomrule
\end{tabular}
\end{table}

\textit{Mining Infrastructures and Code Corpora.} Software Heritage~\cite{dicosmo2017swh,dicosmo2020referencing}, the SWH Graph~\cite{pietri2019swhgraph,boldi2020swhgraph}, and World of Code~\cite{ma2019woc} link content-addressed blobs, commits, and projects at scale. Recent datasets attach licenses~\cite{gonzalezbarahona2023swhlicense}, copy-based reuse~\cite{jahanshahi2024copybased}, and provenance~\cite{jahanshahi2025cracks} to blob IDs, while D\'ej\`aVu~\cite{lopes2017dejavu}, CodeSearchNet~\cite{husain2019codesearchnet}, and The Vault~\cite{nguyen2023vault} redistributed GitHub functions or duplicates. \tool{} adds a source-free structural layer (metrics, byte spans, and shapes) keyed by \code{blob\_id}.

\textit{AST Indexing and Clone Detection.} Boa~\cite{dyer2013boa}, CodeQL~\cite{avgustinov2016ql}, and Glean~\cite{glean} query custom databases built from source, while Comby~\cite{vantonder2019comby} and \code{ast-grep}~\cite{astgrep} scan local disks. HyperAST~\cite{ledilavrec2022hyperast,ledilavrec2025hyperast} hash-conses ASTs across versions of a single repository, where cross-commit sharing is high, unlike independent files. \tool{} combines AST-hashing clone detection~\cite{baxter1998clone,jiang2007deckard,sajnani2016sourcerercc} (exposing training-set duplication~\cite{allamanis2019adverse,lee2022dedup}) with feature indexing~\cite{yan2004gindex,zhang2006fix} for candidate pruning.

\section{Limitations, Governance, and Availability}
\label{sec:limits}

\textit{Limitations and Future Work.} Our public release covers a 40\% uniform sample across ten of fifteen Stack-Edu languages, targets file- and function-level curation and indexing rather than cross-file analysis, and matches Type-1/2 structural clones rather than near-miss (Type-3) edits. As future work, we plan to: (i)~port ingestion to Rust to index the remaining 60\% of Stack-Edu ($\sim$11\,h) and scale to $10^8$+ files across The Stack~v2; (ii)~deploy a server-side filtering API returning \code{view.json} manifests without local index downloads; and (iii)~join \code{shape} clusters with the SWH provenance graph to study cross-repository reuse and AST deduplication for LLM training.

\textit{Governance and Availability.} \tool{} distributes no source text: only upstream metadata, syntactic counts, byte spans, and 64-bit hashes (indexed literals are hashed, though common tokens are dictionary-guessable). We add no author/commit data; blobs were streamed with SWH's agreement, users fetch under upstream terms, and future releases drop withdrawn IDs. Code (MIT): \url{https://github.com/halykoss/swhforest}; data (CC-BY-4.0): \url{https://huggingface.co/datasets/halykoss/swhforest} (Zenodo DOI pending).

\section*{Acknowledgments}
This work was made possible by Software Heritage, the universal source code archive: \url{https://www.softwareheritage.org}. Paolo Ferragina was supported by the Alfred P.~Sloan Foundation with the grant \#\href{https://sloan.org/grant-detail/g-2025-25193}{G-2025-25193}.

\newpage
\let\oldthebibliography\thebibliography
\renewcommand{\thebibliography}[1]{\oldthebibliography{#1}\setlength{\itemsep}{-1.6pt}}
\bibliographystyle{IEEEtran}
\bibliography{refs}

\end{document}

%% file: numbers.tex
\newcommand{\NRelBlobs}{52.1}
\newcommand{\NRelBlobsExact}{52,145,222}
\newcommand{\NRelDefs}{341}
\newcommand{\NRelReposM}{16.8}

\newcommand{\NRelTabGB}{17.9}
\newcommand{\NRelIdxGB}{19.8}
\newcommand{\NRelMissing}{84}
\newcommand{\NRelHours}{7.1}
\newcommand{\NRelRate}{2,000}
\newcommand{\NRelSelPct}{39.95--40.03}

\newcommand{\NRelPyBlobs}{9.48}
\newcommand{\NRelPyDefs}{43.4}
\newcommand{\NRelPyDoc}{23}
\newcommand{\NRelPyErr}{1.4}
\newcommand{\NRelPyTab}{2.82}
\newcommand{\NRelPyIdx}{3.16}

\newcommand{\NRelJsBlobs}{5.17}
\newcommand{\NRelJsDefs}{39.5}
\newcommand{\NRelJsDoc}{13}
\newcommand{\NRelJsErr}{2.1}
\newcommand{\NRelJsTab}{2.04}
\newcommand{\NRelJsIdx}{1.77}

\newcommand{\NRelJavaBlobs}{17.43}
\newcommand{\NRelJavaDefs}{135.6}
\newcommand{\NRelJavaDoc}{41}
\newcommand{\NRelJavaErr}{0.7}
\newcommand{\NRelJavaTab}{6.15}
\newcommand{\NRelJavaIdx}{7.52}
\newcommand{\NRelJavaBoiler}{33.8}
\newcommand{\NRelCppBlobs}{6.45}
\newcommand{\NRelCppDefs}{40.0}
\newcommand{\NRelCppDoc}{15}
\newcommand{\NRelCppErr}{9.6}
\newcommand{\NRelCppTab}{2.25}
\newcommand{\NRelCppIdx}{2.71}

\newcommand{\NRelCsBlobs}{4.56}
\newcommand{\NRelCsDefs}{30.5}
\newcommand{\NRelCsDoc}{24}
\newcommand{\NRelCsErr}{1.3}
\newcommand{\NRelCsTab}{1.65}
\newcommand{\NRelCsIdx}{1.81}

\newcommand{\NRelPhpBlobs}{3.90}
\newcommand{\NRelPhpDefs}{20.8}
\newcommand{\NRelPhpDoc}{50}
\newcommand{\NRelPhpErr}{1.1}
\newcommand{\NRelPhpTab}{1.17}
\newcommand{\NRelPhpIdx}{0.86}
\newcommand{\NRelPhpBoiler}{24.9}
\newcommand{\NRelCBlobs}{2.27}
\newcommand{\NRelCDefs}{8.4}
\newcommand{\NRelCDoc}{15}
\newcommand{\NRelCErr}{12.1}
\newcommand{\NRelCTab}{0.64}
\newcommand{\NRelCIdx}{0.74}
\newcommand{\NRelCBoiler}{4.4}
\newcommand{\NRelTsBlobs}{1.67}
\newcommand{\NRelTsDefs}{12.5}
\newcommand{\NRelTsDoc}{24}
\newcommand{\NRelTsErr}{0.5}
\newcommand{\NRelTsTab}{0.66}
\newcommand{\NRelTsIdx}{0.66}

\newcommand{\NRelGoBlobs}{0.75}
\newcommand{\NRelGoDefs}{4.8}

\newcommand{\NRelGoErr}{0.4}
\newcommand{\NRelGoTab}{0.26}
\newcommand{\NRelGoIdx}{0.26}

\newcommand{\NRelRustBlobs}{0.45}
\newcommand{\NRelRustDefs}{5.1}
\newcommand{\NRelRustDoc}{35}
\newcommand{\NRelRustErr}{2.9}
\newcommand{\NRelRustTab}{0.23}
\newcommand{\NRelRustIdx}{0.29}

\newcommand{\NRelCurMs}{38}
\newcommand{\NRelCurMatches}{117,203}

\newcommand{\NRelCurSaved}{72}

\newcommand{\NRGeoSpeed}{21}
\newcommand{\NRMinSpeed}{3.1}

\newcommand{\NRGeoSpeedLoad}{7}

\newcommand{\NTokRatio}{4.3}
\newcommand{\NEExceptPassCand}{207,111}
\newcommand{\NEExceptPassTotal}{<0.01}

\newcommand{\NETrainFitCand}{9,109}
\newcommand{\NETrainFitTotal}{28.0}
\newcommand{\NETrainFitHits}{7,568}
\newcommand{\NETrainFitFetch}{16}

\newcommand{\NRelIdxLoadS}{104}
\newcommand{\NSgN}{32}
\newcommand{\NSgExact}{5}
\newcommand{\NSgFetchPct}{0.3}
\newcommand{\NExactPct}{27}
\newcommand{\NHumanEvalHits}{15}
\newcommand{\NMbppHits}{227}
\newcommand{\NValChecked}{44,361}
\newcommand{\NValAgree}{99.98}
\newcommand{\NValSkipPct}{11}

\newcommand{\NForestRatio}{4.1}
\newcommand{\NRelCurAllMs}{63}
\newcommand{\NRelCurAllMatchesM}{1.26}
\newcommand{\NRelCurAllGB}{25.7}
\newcommand{\NRelSizeOnlyAllM}{2.42}
\newcommand{\NRelSizeOnlyAllGB}{47.1}
\newcommand{\NRelCurAllSaved}{46}
\newcommand{\NMLSoundN}{61}
\newcommand{\NMLCandMedPct}{0.53}

\newcommand{\NMLExactN}{16}
\newcommand{\NMLGeoSpeed}{52}
\newcommand{\NMLMaxSpeed}{2,032}

\newcommand{\NECForReallocCand}{25,157}
\newcommand{\NECForReallocTotal}{76.8}
\newcommand{\NECForReallocHits}{5,014}
\newcommand{\NECForReallocFetch}{41}

\newcommand{\NECppRangeEmplaceCand}{34,180}
\newcommand{\NECppRangeEmplaceTotal}{132.1}
\newcommand{\NECppRangeEmplaceHits}{13,441}
\newcommand{\NECppRangeEmplaceFetch}{87}

\newcommand{\NECsUsingReadCand}{25,397}
\newcommand{\NECsUsingReadTotal}{68.8}
\newcommand{\NECsUsingReadHits}{22,881}
\newcommand{\NECsUsingReadFetch}{41}

\newcommand{\NEGoDeferUnlockCand}{20,625}
\newcommand{\NEGoDeferUnlockTotal}{94.6}
\newcommand{\NEGoDeferUnlockHits}{17,594}
\newcommand{\NEGoDeferUnlockFetch}{77}
\newcommand{\NEJavaFornameCand}{149,572}
\newcommand{\NEJavaFornameTotal}{0.52}
\newcommand{\NEJavaSyncWaitCand}{19,535}
\newcommand{\NEJavaSyncWaitTotal}{50.8}
\newcommand{\NEJavaSyncWaitHits}{18,476}
\newcommand{\NEJavaSyncWaitFetch}{32}

\newcommand{\NEPhpForCurlCand}{6,355}
\newcommand{\NEPhpForCurlTotal}{28.5}
\newcommand{\NEPhpForCurlHits}{523}
\newcommand{\NEPhpForCurlFetch}{11}

\newcommand{\NERustUnsafeTransmuteCand}{3,321}
\newcommand{\NERustUnsafeTransmuteTotal}{11.1}
\newcommand{\NERustUnsafeTransmuteHits}{2,959}
\newcommand{\NERustUnsafeTransmuteFetch}{6}

\newcommand{\NETsTryConsoleErrCand}{17,174}
\newcommand{\NETsTryConsoleErrTotal}{38.8}
\newcommand{\NETsTryConsoleErrHits}{14,400}
\newcommand{\NETsTryConsoleErrFetch}{27}
\newcommand{\NEJsAwaitInLoopCand}{15,959}
\newcommand{\NEJsAwaitInLoopTotal}{<0.01}